\documentclass[normal]{aa}
\usepackage{psfig}

\def\s4{S4 0954+658}
\def\etal{et al.}

\begin{document}
\title{Multi-band optical micro-variability observations of the BL Lac 
object S4 0954+658}
\author{I.E. Papadakis\inst{1} \and  V. Samaritakis\inst{1} \and P. Boumis 
\inst{2} \and J. Papamastorakis\inst{1,3} }
\offprints{I. E. Papadakis;  e-mail: jhep@physics.uoc.gr}
\institute{Physics Department, University of Crete, P.O. Box 2208,
   710 03 Heraklion, Greece
\and Institute of Astronomy \&
Astrophysics, National Observatory of Athens, I. Metaxa \& V. Paulou, P.
Penteli, GR-15236, Athens, Greece
\and IESL, Foundation for Research and Technology-Hellas, P.O.Box
1527, 711 10 Heraklion, Crete, Greece}
\date{Received 16 March 2004 / Accepted 30 June 2004}
\abstract{We have observed \s4\ in the $B, V, R$ and $I$ bands for one
night in March and two nights in April, 2001, and in the $B$ and $I$ bands
for four nights in May, 2002.  The observations resulted in almost evenly
sampled light curves, $3-8$ hours long, with an average sampling interval
of $\sim 5-15$ min. Because of the dense sampling and the availability of
light curves in more than one optical bands we are able to study the
intra-night flux and spectral variability of the source in detail.
Significant observations were observed in all but one cases. On average,
the flux variability amplitude, on time scales of minutes/hours, increases
from $\sim 2-3\%$ in the $I$, to $\sim 3.5-5.5\%$ in the $B$ band light
curves. We do not detect any flares within the individual light curves.
However, there is a possibility that the April 2001 and late May 2002
observations sample two flares which lasted longer than $\sim 1-3$ days.
The evidence is only suggestive though, due to the limited length of the
present light curves with regard to the duration of the assumed flares. No
spectral variations are detected during the April 2001 observations. The
source flux rises and decays with the same rate, in all bands.  This
variability behaviour is typical of \s4, and is attributed to geometrical
effects. However, significant spectral variations are observed in May
2002. We find that the spectrum hardens/softens as the flux
increases/decreases, respectively. Furthermore, the ``hardening" rate of
the energy spectrum is faster than the rate with which the spectrum
becomes ``redder" as the flux decays. We also find evidence (although of
low statistical significance) that the $I$ band variations are delayed
with respect to the $B$ band variations. If the May 2002 observations
sample a flaring event, these results suggest that the variations are
caused by energetic processes which are associated with the particle
cooling and the source light travel time scales.
}

\titlerunning{Optical micro-variability of S4 0954+658}
\authorrunning{Papadakis \etal}
\maketitle
   
\section{Introduction}
\smallskip

BL Lac objects are one of the most peculiar classes of active galactic
nuclei (AGN). They show high polarization (up to a few percent, as opposed
to less than $\sim 1\%$ for most AGNs) and usually do not exhibit strong
emission or absorption lines in their spectra. They also show continuum
variability at all wavelengths at which they have been observed, from
X-rays to radio. In the optical band they show large amplitude, short time
scale variations. The overall spectral energy distribution of BL Lacs
shows two distinct components in the $\nu-\nu F_{\nu}$ representation. The
first one peaks from mm to the X--rays, while the second component peaks
at GeV--TeV energies (e.g. Sambruna \etal\ 1996). The commonly accepted
scenario assumes that the non-thermal emission from BL Lacs is synchrotron
and inverse-Compton radiation produced by relativistic electrons in a jet
oriented close to the line of sight.

\s4\ is a well studied BL Lac object in the optical bands. Its optical
variability has been studied by Wagner \etal\ (1990, 1993). Wagner \etal\
(1990) found large amplitude variations (of the order of $\sim 100\%$) on
time scales as short as 1 day. Wagner \etal\ (1993), using a well sampled,
1 month long $R$ band light curve, observed symmetric flares, with a
max--to--min variability amplitude of the order of $\sim 4-5$. Raiteri
\etal\ (1999) presented a comprehensive study of the optical (and radio)
band variability of the source using $4$ year long optical light curves.
They also detected large amplitude, fast variations. Studying the $B-R$
colour variations, they also found that the mid and long-term variations
in the source are not associated with spectral variations.

In this work, we present simultaneous $B$, $V$, $R$ and $I$ band
monitoring observations of \s4, which were obtained in the years 2001 and
2002 from Skinakas Observatory, Crete, Greece. The quality of the light
curves is similar to those presented by Papadakis \etal\ (2003) in the
case of BL Lac itself. Compared to previous observations, the biggest
advantage of the present observations is that they resulted in light
curves at different energy bands, with a dense, almost evenly sampling
pattern.

Our main aim is to use these light curves in order to investigate the flux
and spectral variations of the source on time scales as short as a few
minutes/hours.  We find that, apart from variations which are best
explained as being the product of geometric variations (in agreement with
the past observations), the source also exhibits fast variations which are
most probably caused by changes in the energy distribution of the high
energy particles in the synchrotron emitting source.

\section{Observations and data reduction} 

\s4\ was observed for 3 nights in 2001 and 4 nights in 2002 with the 1.3
m, f/7.7 Ritchey-Cretien telescope at Skinakas Observatory in Crete,
Greece. The observations were carried out through the standard Johnson $B,
V$ and Cousins $R, I$ filters. The CCD used was a $1024 \times 1024$ SITe
chip with a 24 $\mu$m$^{2}$ pixel size (corresponding to
$0^{\prime\prime}.5$ on the sky). The exposure time was 300, 240, 120 and
120 s for the $B, V, R$ and $I$ filters, respectively. In Table 1 we list
the observation dates, and the number of frames that we obtained each
night. During the observations, the seeing varied between $\sim
1^{\prime\prime} - 2^{\prime\prime}$. Standard image processing (bias
subtraction and flat fielding using twilight-sky exposures) was applied to
all frames.

\begin{table}
\begin{center}
\caption{Date of observations, and number of frames (nof) obtained at
each optical band.}
\begin{tabular}{lcccc} \hline
Date & $B$ & $V$ & $R$ & $I$ \\
     & (nof) & (nof) & (nof) & (nof) \\
\hline
26/03/01 & -- & -- & 24 & 24 \\
18/04/01 & 16 & 16 & 17 & 16 \\
19/04/01 & 5  & 5  & 5  & 5  \\
23/05/02 & 27 & -- & -- & 27 \\
28/05/02 & 19 & -- & -- & 19 \\
29/05/02 & 24 & -- & -- & 24 \\
20/05/02 & 24 & -- & -- & 24 \\
\hline
\end{tabular}
\end{center}
\end{table}
 
We performed aperture photometry of \s4\ and of the comparison stars 2, 3,
4, and 7 of Raiteri \etal\ (1999) by integrating counts within a circular
aperture of radius $10^{\prime\prime}$ centered on the objects. The
remaining 5 comparison stars in the list of Raiteri \etal\ are also
visible in the field of view. However the scatter in their light curves
was larger than the scatter in the light curves of the 4 comparison stars
that we kept. For that reason we decided not to use them. The error
on the \s4\ magnitude measurements in each frame were estimated as
follows. First, we estimated the standard deviation ($\sigma$) of the
comparison stars light curves, in each band, during the 2001 and 2002
observations. We consider these values as representative of the
uncertainty in the magnitude estimation of the comparison stars within
each observing period. Then, for each frame, the error on the \s4\
magnitude was computed using the standard ``propagation of errors" formula
of Bevington (1969), taking into account the photometry error of the
source's measurement and the $\sigma$ values of the comparison stars.

The calibrated magnitudes were corrected for reddening using the
relationship $A_{V}=N_{H}/2.23\times10^{21}$ (Ryter 1996), in order to
estimate the $V$ band extinction ($A_{V}$, in magnitudes). We assumed the
column density value of $N_{H}=9.0\times 10^{20}$ cm$^{-2}$, derived from
X--ray measurements with ROSAT (Urry \etal\ 1996). This value implies
$A_{V}=0.40$, which is almost identical to the respective value of
Schlegel \etal\ (1998) as taken from NED\footnote{The NASA/IPAC
Extragalactic Database (NED) is operated by the Jet Propulsion Laboratory,
California Institute of Technology, under contract with the National
Aeronautics and Space Administration.}. Using the $A_{\lambda}$ versus
$\lambda$ relationship of Cardelli \etal\ (1989) we found the extinction
(in magnitudes) in the $B, R,$ and $I$ filters: $A_{B}=0.53, A_{R}=0.30$,
and $A_{I}=0.19$. Finally, we converted the dereddened magnitudes into
flux, without applying any correction for the contribution of the host
galaxy, as it cannot be resolved even in deep HST images (Scarpa \etal\
2000).

\section{The observed light curves} 

\begin{figure}
\psfig{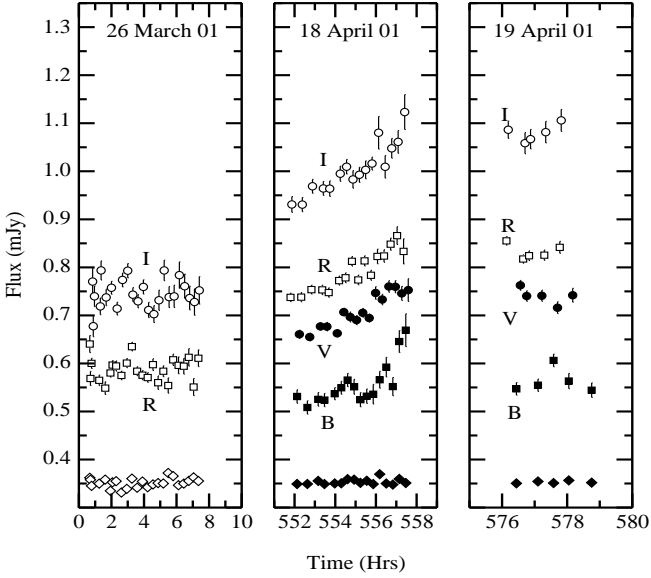}
\caption[]{$B, V, R, $ and $I$ band light curves during the 2001
observations. Time is measured in hours from 19:00 UT on March 26, 2001.
The small filled diamonds below the $B$ band light curve of \s4\ show the
$B$ band light curve of the comparison star 2 (open diamonds in the left
panel show the $R$ band light curve of the same star). For clarity
reasons, the standard star light curves are shifted to the same level of
$\sim$ 0.35 mJy.}
\end{figure}

\begin{figure}
\psfig{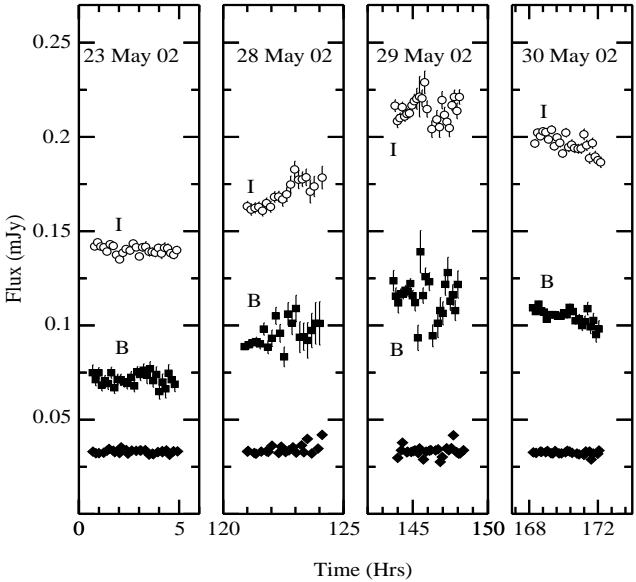}
\caption[]{$B$ and $I$ band light curves during the 2002 observations.
Time is measured in hours from 18:00 UT on May 23, 2002. As before with
Figure~1, the small filled diamonds show the $B$ band light curve of
the comparison star 2 (shifted to a mean of $\sim$ 0.033 mJy).}
\end{figure}

Figs.~1 and 2 show the dereddened $B,V, R,$ and $I$ light curves of \s4\
during the 2001 and 2002 observations, respectively. In the same figures
we also show the $B$ band light curve of the comparison star 2. While the
light curve of this star (and of the other 3 comparison stars in all
bands) does not show significant variations, intrinsic intra-night
variations can be observed in most of the \s4\ light curves.  Use of the
$\chi^{2}$ test shows that the hypothesis of a constant flux can be
rejected at the $99.9\%$ significance level in all cases, except from the
May 23, 2002 $B$ and $I$ light curves.

The source was at a brighter flux state in 2001.  During the April 18
observations the source flux increases. In the same night, the $B$ band
light curve shows a fast flux decrease (at around 557 hrs) which is absent
in the $V,R,$ and $I$ light curves. However, when compared with the
adjacent points, the flux decrease is significant at only the $\sim
2\sigma$ level. The April 19, 2001 light curves last for less than $\sim
3$ hours (due to bad weather we could not observe the source for a larger
period). The $R$ and $I$ band light curves are quite similar, but the $V$
and $B$ band light curves look different. There is a hint that they may be
anti-correlated, but no firm conclusions can be drawn due to the short
duration of the observations. During the May 28, 2002 observations the
source flux increases, while the opposite behaviour is observed in May 30,
2002. In fact, as Fig.~2 suggests, the late May 2002 light curves could be
part of a ``flare-like" event which lasted for more than $\sim$ 3 days. It
is possible that the May 28, and the May 29-30 observations sample the
flux increasing and decaying parts of the flare, respectively.  No trends
on time scales of $\sim$ hours are evident in the $R$ and $I$ band, March
26, 2001 light curves. However, as the $\chi^{2}$ test indicates, the
small amplitude variations around the mean flux level are significant in
this case. We find that $\chi^{2}=84.2$ and $60.8$ for the $R$ and
$I$ band light curves, respectively. The number of degrees of freedom is
23. Consequently, the probability that we would observe these values by
chance, if the flux remained constant during the observations, is less
than $\sim 3\times 10^{-5}$, in both cases. Finally, the peak
max--to--min variability amplitude is $\sim 15\%$ and $\sim 30\%$ during
the April 2001, and May 2002 observations, respectively.

In order to compare the amplitude of the variations that we observe in
the various light curves we computed their ``fractional variability
amplitude" ($f_{rms}$), as described in Papadakis \etal\ (2003). The
average $f_{rms}$ of the 2000 $B$ and $I$ band light curves is
$f_{rms,B}=3.2\pm0.5\%$ and $f_{rms,I}=2.6\pm0.3\%$, respectively (we
have not considered the May 23 light curves, as they show no
significant variations). The average $I$ band variability amplitude
during the 2001 observations is $f_{rms,I}=2.8\pm0.5\%$, consistent
with the 2002 estimate. We also find that the average variability of
the 2001 $R$ band light curves is $f_{rms,R}=4.0\pm0.5\%$, while
$f_{rms}\sim 5\%$ and $5.5\%$ for the April 18, 2001 $V$ and $B$ band
light curves. We conclude that \s4\ shows low amplitude variations
($\sim 2-5\%$ on average) on time scales of a few hours. Furthermore,
the variability amplitude increases towards higher frequencies,
i.e. from the $I$ to the $B$ band light curves.

\begin{figure}
\psfig{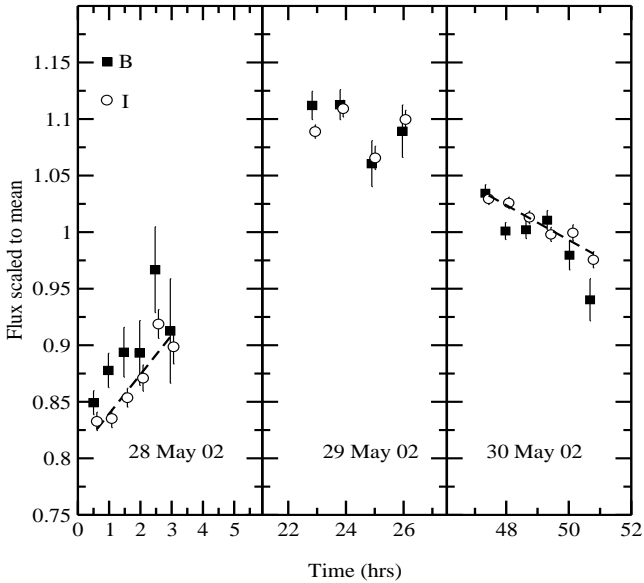}
\caption[]{$B$ and $I$ band light curves in 2002 (filled squares and open
circles, respectively) normalized to their average flux level. Time is
measured in hours from the beginning (plus half an hour) of the May 28 
$B$ band observations. In order to reduce the experimental noise, we
have smoothed the May 28, 29 and 30 light curves using a bin size of 3, 6,
and 4, respectively.}
\end{figure}

\begin{figure}
\psfig{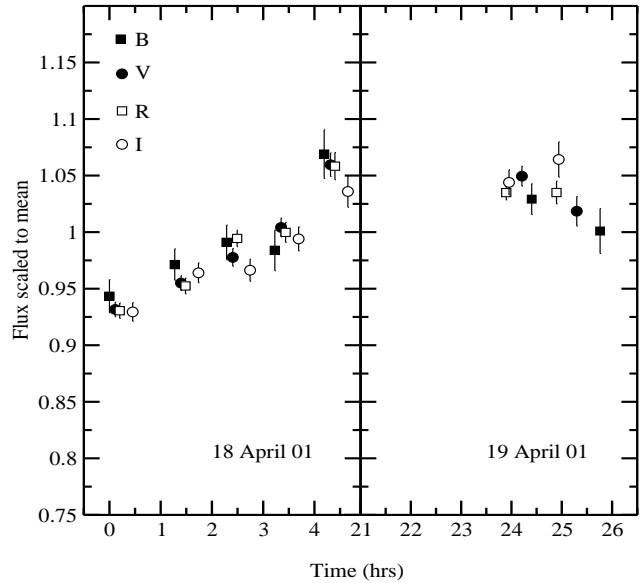}
\caption[]{ Same as Fig.~3 for the April, 2001 light curves (filled
squares, filled circles, open squares, and open circles show the $B, V,
R,$ and $I$ band light curves, respectively). The light curves have been
smoothed out using a bin size of 3, except from the last point where a bin
size of 2 was used. Time is measured in hours from the start of the April
18, $B$ band observations.}
\end{figure}

Figs.~3 and 4 show the late May 2002 and April 2001 light curves,
normalized to their mean. In this way we can compare directly light curves
in different energy bands, and investigate whether there are any
differences in their variability pattern.  On average, the $B$ over
$I$ band flux ratio in May 28 appears to be larger than the same ratio of
the May 29 and 30 observations. To investigate this issue further, we
divided the $B$ and $I$ band normalized fluxes, within each night, and
computed their weighted mean values, $<B/I>$. Our results are as follows:
$<B/I>_{\rm May 28}=1.04\pm0.01, <B/I>_{\rm May 29}= 1.007\pm 0.008,$ and
$<B/I>_{\rm May 30}=0.992\pm0.004$. They imply spectral variations, with
the energy spectrum becoming progressively ``redder" from the May 28 to
the May 30 observations. 
 
Furthermore, Fig.~3 shows clearly that the flux rises steeper than it
decays, in both energy bands. In order to quantify this effect, we fitted
the $I$ band light curves of the May 28 and 30 observations with a linear
model of the form:  $I_{norm}=a+\beta t$. The slope of this line, in units
of ``per cent/hrs", is a measure of the normalized flux variation per unit
time.  The dashed lines in Fig.~3 show the best fitting models, which
describe rather well the observed trends in the $I$ band light curves. The
best fitting slopes are $\beta=3.5\%\pm 0.7\%$ and $-1.5\%\pm0.2\%$
hrs$^{-1}$, for the May 28 and May 30 light curves, respectively. In other
words, while the flux increases with a rate of $\sim 3.5\%$ per hour in
May 28, it decreases with a slower rate (of $\sim 1.5\%$ per hour) in May
30.  These differences suggest the presence of spectral variations during
the May 2002 observations.

The situation is markedly different in April, 2001.  First of all, the
normalized flux rises with the same rate in all bands during the April 18
observation. Indeed, when we fit all the light curves with a linear model,
the best fitting slope suggests a flux increase rate of $\sim 2.5\%$ per
hour in all cases. The April 19 normalized light curves in Fig.~4 appear
to be in agreement with each other. If we consider them together as a
``single" light curve, and we fit them with the same linear model as
above, we find that the flux decreases with a rate of $\sim 2\%$ per hour,
similar to the flux increase rate of the previous night. Fig.~4 suggests
that the April 18 and 19, 2001 light curves could be part of a single
flare (like the late May 2002 observations) which lasted more than a day.
This is only suggestive of course, due to the short duration of the
observations (specially in April 19). However if it is true, our results
imply that during the flare evolution, the flux rises and decays with the
same rate, opposite to what we observe in May 2002.

\section{Spectral variability}

Using the dereddened light curves and the equation,
$\alpha_{12}=\log(F_{1}/F_{2})/\log(\nu_{1}/\nu_{2})$, where $F_{1}$ and
$F_{2}$ are the flux densities at frequencies $\nu_{1}$ and $\nu_{2}$,
respectively, we calculated two-point spectral indices: $B-I$ for the 2001
and 2002 observations, and $V-I$, $R-I$ for the 2001 observations only. We
have used consecutive observations, with a time difference less than $\sim
15$ min.

Fig.~5 shows the $\alpha_{BI}$ versus the $B+I$ normalized flux plot for
the 2002 observations. We observe significant $\alpha_{BI}$ variations
which are correlated with the source flux. During the May 28 observations
the spectrum becomes progressively ``bluer"/harder (i.e. $\alpha_{BI}$
increases) as the source brightens. We observe the same trend during the
May 29 and 30 observations. As the source flux decreases, the spectrum
becomes ``redder"/softer.  However, the spectral variations do not follow
the same ``path" in the [$(B+I)_{norm}, \alpha_{BI}$] plane during the
flux rise and decay phases.  First of all, the average $\alpha_{BI}$ index
in May 28 is larger than the average index during the May 29, and 30
observations. When the flux increases the spectrum is ``bluer" than the
spectrum during the flux decay, in agreement with our comments in the
previous section regarding the normalized light curves in Fig.~3.

We used a simple linear model of the form $\alpha = A+B\times flux$ to
describe quantitatively the relation between flux and spectral index. The
dotted lines in Fig.~5 show the best fitting model to the May 28 and to
the combined May29/30 data.  The model describes the spectral variations
well. The best fitting slopes are $1.0\pm0.2$ and $0.5\pm0.1$ for the
spectral variations during the flux rise and decay phases, respectively.  
Their difference is $0.5\pm0.2$. Although this is only a $2.5 \sigma$
effect, it suggests that the spectral variability rate during the rise and
decay phases of the light curves is different. The rate with which the
spectrum hardens during the flux rise is faster than the rate with which
the spectrum softens during the flux decay phase.

Finally, the arrows in Fig.~5 show the flux evolution during the late May
2002 observations. The source flux increases in May 28, and the spectrum
becomes ``bluer". In the following two nights, the source flux decreases,
and the spectrum becomes ``redder", but at smaller rate. If the late
May 2002 observations are part of the same flaring event, Fig.~5 suggests
that the variation of $\alpha_{BI}$ as a function of the normalized flux
follows a loop-like path, in the clockwise direction, during the flux
rising and decaying parts of the event.

Fig.~6 shows the $\alpha_{BI}$ and $\alpha_{VI}$ versus the respective
normalized flux [i.e. $(B+I_{norm})$ and $(V+I_{norm})$, respectively] for
the April 2001 observations. This figure shows clearly that there are no
spectral variations during these observations. Comparison between the
plots in Figs.~5 and 6 implies that the source varies in a different way
during the 2001 and 2002 observations. This is in agreement with the
result we reached in the previous section, when we compared the normalized
2001 and 2002 light curves. The fact that during the April 18 observations
the spectral indexes $\alpha_{BI}, \alpha_{VI}$, and $\alpha_{RI}$ (not
shown in Fig.~6 for clarity reasons) remain constant is consistent with
the result that the flux rise rate is the same at all energy bands. The
spectral indices remained constant in the following night as well. If
the April 18 and 19 observations are part of the same flare, the flux
rise and decay rate are equal in all optical bands. We conclude that the
April 18 2001 flare is symmetric, and evolves with no spectral variations.

\begin{figure}
\psfig{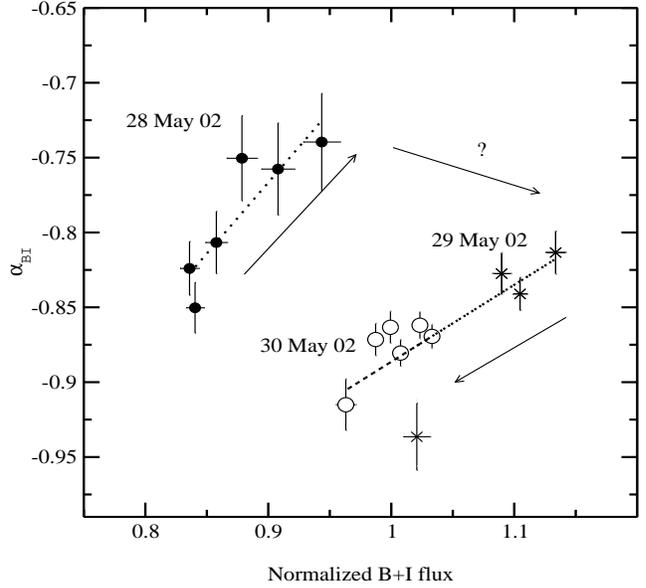}
\caption[]{$B-I$ spectral index vs. the $B+I$ normalized flux for the 2002
observations. For clarity reasons, we have smoothed the data using the
same bin size that we have used for the respective light curves in Fig.~3.  
The dotted lines show the best linear model fit to the data, while the
arrows indicate the flux variability trend (i.e. flux increase or
decrease)  during the observations (see text for details).}
\end{figure}

\begin{figure}
\psfig{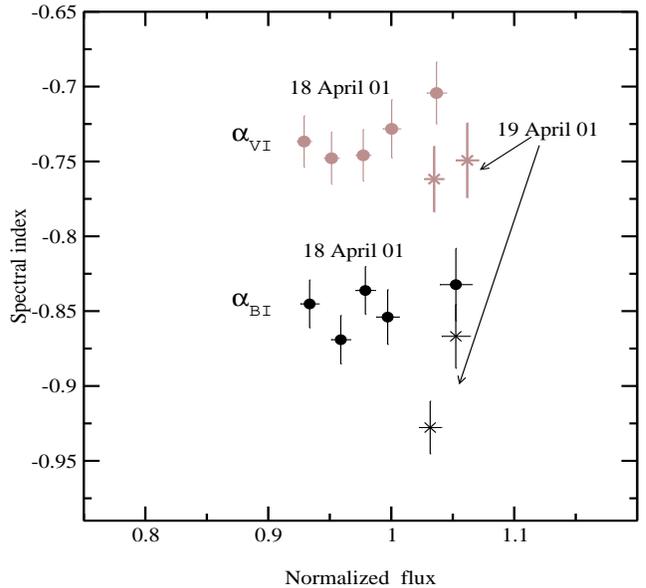}
\caption[]{$B-I$ (filled black circles and crosses) and $V-I$ (grey
filled circles and crosses) spectral indices vs the $B+I$ and $V+I$
normalized source flux for the 2001 observations. The data have been
smoothed out using the same bin size as with the respective light curves 
in Fig.~4.}
\end{figure}

\section{Discussion and conclusions}

We have observed \s4\ in four optical bands, namely $B$, $V$, $R$, and
$I$, for 2 nights in 2001 and in the $B$ and $I$ bands for 4 nights in
2002. Most light curves last for $\sim 4.5$ hours. There are $15-25$
points in each of them, almost evenly spaced, with an average sampling
interval of $\sim 0.1-0.2$ hours, on average. Our results can be
summarized as follows:

1) The source shows intrinsic, low amplitude variations ($\sim 2-5$
percent) on time scales as short as a few hours. In most cases, we observe
flux rising or decaying trends within each light curve. We also observe
source variations around a constant flux level (March 26, 2001). In one
night (May 23, 2002) we do not detect any significant variations.  The
variability amplitude decreases from the $B$ to the $I$ band light curves.

2) On longer time scales, we observe larger amplitude variations (of the
order of $\sim 20-30\%$) within $1-2$ days, while the average source flux
decreased by a factor of $\sim 5$ between the 2001 and 2002 observations.
This variability pattern is very similar to what has been observed in the
past from the same source, on both long and short time scales (e.g. Raiteri
\etal\ 1999; Wagner \etal\ 1993).

3) We do not observe spectral variations in April 2001. On the contrary,
the May 2002 observations show significant spectral variations which are
correlated with the source flux. The spectrum becomes ``bluer"/``redder"  
as the flux increases/decreases, respectively. The spectral variability
rate when the flux increases is faster than the variability rate during
the flux decay.

We have not detected any flares within the individual light curves.
However, Figs.~2,3, and 4 suggest that the April 2001 and late May 2002
light curves {\it could} be part of two flares which lasted for more than
$1-3$ days. We cannot be certain of course, due to the short duration of
the present light curves. However, if that were the case, the results
listed above could impose interesting constrains on the mechanism which
causes the observed variations in the source. For that reason, we
investigate below briefly the consequences of our results, {\it assuming}
that the April 2001 and May 2002 observations presented in this work are
indeed part of two flares, which lasted for a few days.

Wagner \etal\ (1993) detected several flares in the course of an $R$ band
four week monitoring of \s4. In all cases, the flares were symmetric, with
equally fast rising and declining parts. These authors suggested a
geometric explanation for these variability events. If the optical
emission is produced in discrete blobs moving along magnetic fields, and
if the viewing angle $\theta$ (i.e. the angle between the blob velocity
vector and the line of sight) varies with time, then the beaming or
Doppler factor should vary accordingly. Since the observed flux depends on
this, symmetric $\theta$ variations should result in symmetric flares as
well. Although our observations have not fully resolved such an event, the
April 2001 observations, which show equally fast rising and decaying flux
changes, support the hypothesis that we are witnessing such a flare.

The big advantage of the present observations is the availability of light
curves in four optical bands. If the observed variations are caused by
variations of the Doppler factor, we would expect them to be ``achromatic"
(assuming no spectral break within the considered optical bands), like the
April 2001 observations which show no spectral changes during the flux
variations. A change of less than 0.5 degrees only can explain flux
variations of the order of $20\%$ (Ghisellini \etal\ 1998), similar to
what we observe during the April 2001 observations. Raiteri \etal\ (1999)
found that the long term optical variations of \s4\ are also achromatic.
Obviously, changes of the jet orientation with respect to the observer's
line of sight is a major cause for the observed long and short term
variations in this source.

However, the May 2002 observations indicate that some of the variations
seen in the light curves of this source do {\it not} have geometric
origin, as we find significant energy spectral variations which are
correlated with the flux variations. Energetic processes associated with
the particle acceleration and cooling (i.e. Kirk, Rieger, \& Mastichiadis,
1998; Chiaberge \& Ghisellini, 1999) are probably necessary in order to
explain them. For example, in May 2002 we observe the spectrum becoming
increasingly bluer during the flux rising phase (as the $\alpha_{BI}$ vs
flux plot in Fig.~5 shows). This suggests that the $B$ band light curve
rises faster than the $I$ band light curve. In this case the acceleration
process of the energetic particles cannot dominate the observed
variations. Since the acceleration time scale is shorter for lower energy
particles, we would expect to observe the opposite trend in this case,
i.e. the lower energy light curves to show steeper rising phases. The
observed spectral variability can be explained if the injection of the
high energy particles is instantaneous, and the volume of the $B$ band
emitting source increases faster than the cooling time scale. In this case
it takes longer time for the higher energy particles to cool and start
emitting at lower frequencies, than the time it takes for the B band
emitting volume, and hence the flux as well, to increase. Furthermore, as
the cooling time scale is faster for the higher energy particles, we
expect the decaying part of the flare to be steeper in the $B$ band light
curve as the flux decreases, and the spectrum to become redder, exactly as
observed.

\begin{figure}
\psfig{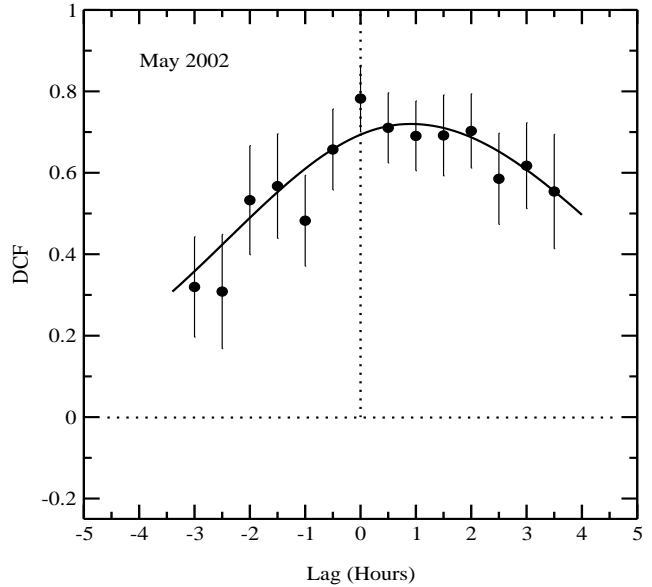}
\caption[]{The Discrete Correlation Function between the $B$ and $I$ band
light curves during the late May 2002 observations. Significant
correlation at positive lags means that the $B$ band variations are
leading the $I$ band.  The solid line shows the best fitting Gaussian
model to the data.}
\end{figure}

If indeed the observed variability in late May 2002 observations
propagates from higher to lower frequencies, we should expect to observe
delays between the $B$ and $I$ light curves, with the $B$ band variability
leading the $I$ band variations. In order to investigate this issue, we
estimated the discrete correlation function (DCF) of the the $B$ and $I$
light curves, using the May 28, 29 and 30, 2002 data, and the method of
Edelson \& Krolik (1988).  The DCF is shown in Fig.~7.  The maximum DCF
value is $\sim 0.7-0.8$, which indicates that the two light curves are
highly correlated. This is not surprising given the very good agreement
between the two light curves (Fig.~3). In order to quantify the delay
between the light curves (i.e. the lag at which the maximum DCF occurs,
$k_{max}$), we fitted the DCF with a Gaussian (the best fitting model is
shown with the solid line in Fig.~7). The best fitting $k_{max}$ value is
$0.9\pm 1.0$ hrs. The uncertainty corresponds to the $90\%$ confidence
limit, and was estimated using the Monte Carlo techniques of Peterson
\etal\ (1998). This result indicates that there is a delay between the $B$
and $I$ band variations, in the expected direction, although not
statistically significant. Nevertheless, Fig.~7 shows clearly that the DCF
is highly asymmetric towards positive lags. This asymmetry suggests that
the $I$ band light curve is indeed delayed with respect to the $B$ band
light curve, although in a rather complicated way. 

We conclude that, if the May 2002light curves are part of the same
flaring event, the observed variations are probably caused perturbations
which activate the energy distribution of the particles in the jet. The
fact that the rising time scales are steeper in the $B$ band light curves
and the DCF asymmetry towards positive lags imply that the injection time
scales are very short and that the variations are governed by the cooling
time scale of the relativistic particles and the light crossing time
scale.

The variability that we observe in May 2002 is qualitatively similar to
the variability observed in the short term, optical light curves of BL Lac
itself (Papadakis \etal\ 2003). Just like BL Lac, \s4\ is a classical
radio selected BL Lac, whose spectral energy distribution shows a peak at
frequencies below the optical band (e.g. Raiteri \etal\ 1999).  
Therefore, the optical band corresponds to frequencies located above the
peak of the synchrotron emission in this object.  We believe that our
results demonstrate that well sampled, multi-band optical, intra-night
observations of BL Lac objects, whose peak of the emitted power is at
mm/IR wavelengths {\it can} offer us important clues on the acceleration
and cooling mechanism of the particles in the highest energy tail of the
synchrotron component.

The present observations are not long enough in order to define
accurately the characteristic time scales of the variability processes
in action in \s4. Longer, continuous monitoring of the source, with
the collaboration of more than one observatory is needed to this
aim. We hope that the results of this work, will motivate the
collaboration between the various observatories towards this
direction.

\vskip 0.4cm

\begin{acknowledgements}
Skinakas Observatory is a collaborative project of the University of
Crete, the Foundation for Research and Technology-Hellas, and the
Max--Planck--Institut f\"ur extraterrestrische Physik.
\end{acknowledgements}

\end{document}